\documentclass [11pt, a4paper]{article}
\usepackage{vmargin,color}
\setmarginsrb{2.0cm}{2.5cm}{2.0cm}{1.0cm}{0.0cm}{0.0cm}{0.0cm}{1.0cm}
\usepackage[french,english]{babel}
\usepackage[T1]{fontenc}
\usepackage{cite,graphicx,float,setspace,amssymb,amsmath}
\usepackage{graphicx,float,setspace,amssymb,amsmath}

\newcommand{\equa}[1]{\begin{eqnarray} \label{#1}}
\newcommand{\auqe}{\end{eqnarray}}
\newcommand{\equab}[1]{\begin{align}\label{#1}}
\newcommand{\bauqe}{\end{align}}
\newcommand{\tab}[1]{\begin{tabular}{#1}}
\newcommand{\bat}{\end{tabular} \\ }
\newcommand{\blanc}{\makebox[0.5 cm]{ }}

 \newcommand{\be}{\beta} \newcommand{\la}{\lambda}
\newcommand{\ga}{\gamma} \newcommand{\ep}{\epsilon}
\newcommand{\de}{\delta} \newcommand{\gD}{\Delta} 

\providecommand{\abs}[1]{\left\vert#1\right\vert}

 \newcommand{\s}{\sigma} \newcommand{\om}{\omega}
\newcommand{\Om}{\Omega}
\newcommand{\tend}{\rightarrow}
\newcommand{\mt}[1]{\widetilde{#1}}

\newcommand{\ddf}{\frac{\partial^2\phi}{\partial z^2}}

\newcommand{\fik}{\varphi_{k}(z)}
\newcommand{\fix}[1]{\varphi_{#1}(z)}

\newcommand{\szz}{S_{\gD z\;\gD z}}
\newcommand{\hb}{\bar{h}_1}
\newcommand{\zs}{z_{int}(\vec{s})}
\font\bba=msbm10 scaled 1080
\font\bbb=msbm8 %scaled 1080
\font\bbc=msbm6 %scaled 1080
\newfam\bbfam
\textfont\bbfam=\bba
\scriptfont\bbfam=\bbb
\scriptscriptfont\bbfam=\bbc
\def\bb{\fam\bbfam\bba}
\def\R{{\bb R}}

\usepackage{ucs}
\usepackage[utf8x]{inputenc} 
\usepackage{textcomp}
\usepackage{pifont}
\begin{document} 
\selectlanguage{english} 
% \articletype{}
%
\title{
{\bf Statistical field theory for  liquid vapor interface}
}
\author
{~V. Russier$^a$         \footnote{~russier@icmpe.cnrs.fr} 
~and ~J.-M. Caillol$^b$  \footnote{~jean-michel.caillol@th.u-psud.fr}  \\
$^a$~ICMPE, UMR 7182 CNRS and Universit\'e Paris Est \\
2-8 rue Henri Dunant, 94320 Thiais, France. \\
$^b$~Laboratoire de Physique Th\'eorique, UMR 8627  CNRS and 
Universit\'e de Paris Sud, \\ bat. 210, 91405 Orsay Cedex, France. 
}
\date{}
\maketitle
\begin{abstract}
% Resume 
A statistical field theory for an inhomogeneous liquid, a planar
liquid/vapor interface, is devised from first principles. The grand 
canonical partition function is represented {\it via} a
Hubbard-Stratonovitch tranformation leading, close to the critical point,
 to the usual  $\phi^4$ scalar field theory which is then rigorously 
considered at  the one-loop level. 
When further simplified it yields the well-known capillary wave theory 
without any {\it ad hoc} phenomenological parameter. 
Internal coherence of the one-loop approximation is  discussed and good 
overall qualitative agreement with recent numerical simulations is stressed.
\\
PACS: s61.30.Hn; 64.70.F; 68.03.Cd.  
\end{abstract}
%}
\eject
\section{Introduction } 
\label{intro} 
%%%% Introduction apres cette ligne 
The structure in the interface between two fluid phases at coexistence plays a key role in many 
specific situations, such as, for instance, the wetting transition and related phenomena
\cite{sull, lipo_1,dietrich_1,lipo_5, brez, parry_s, hrt}. An understanding and proper
description  of this structure is of course challenging and presents a great interest in itself from a 
theoretical point of view. For simple fluids, one of the key features of interfaces  that separate 
subcritical phases at coexistence is the presence of  thermally activated capillary waves \cite{bls} 
coupled with density fluctuations in the bulk. Since the pioneering work of Buff {\it et al.} \cite{bls} 
capillary waves have been widely studied and are usually described through the introduction of an
\textit{ad hoc} effective surface Hamiltonian, written as a functional of the interface height, $\zs$,
$\vec{s}$ being the coordinate parallel to the interface -in the case of a planar interface as considered
henceforth-.  $H[\zs]$ can be obtained through different ways and is most generally 
deduced from phenomenological arguments taking into account the interplay between gravity and  
surface tension effects. It is noteworthy that most works in this area are devoted to the wetting 
transitions where the Hamiltonian includes an external potential, say $W(\zs)$, which  pins the 
interface to a solid substrate. 

The Hamiltonian $H[\zs]$, can be treated in the framework of statistical field theoretical methods \cite{lipo_1,lipo_5}, including  renormalization group (RG) \cite{wet_rg_1} technics.
For instance, recently, the hierarchical reference theory (HRT) has been generalized to
inhomogeneous cases in order to deal with the wetting transition \cite{hrt}.
In the case of the  liquid/vapor interface, when $H[\zs]$ is treated at the gaussian level,  
the capillary wave theory (CWT) is recovered, leading to the well known $1/q^2$ behavior for 
the height-height structure factor $\szz(q)$. Moreover, $3D$ $\phi^4$ model in which the stochastic 
variable $\phi$ does not coincide  with the height $\zs$ were studied in refs
\ \cite{brez, parry_b, parry_s, brill_surf}.
Alternative ways to deduce an effective Hamiltonian from a density functional theory (DFT)
including the effect of gravity \cite{romero_92} or the local curvature of the interface
\cite{eff_h_dietr,dft_md} have been also considered. Recently \cite{blokhuis}, a model for the 
density profile based on an extension of a displaced profile approximation, \textit{i.e.} in which
the profile is written as a function of $(z-\zs)$, was considered. In addition, the model takes into 
account,  together with the surface fluctuations described by the height $\zs$,  bulk phase 
fluctuations.

The purpose of the present work is to provide a simple description of the \mbox{liquid / vapor} 
interface  of  fluids in the framework of an exact statistical field theory. The latter is obtained 
from a Kac-Siegert-Stratonovich-Hubbard-Edwards (KSSHE)
\cite{Kac,siegert,Strato,Hubbard,Edwards}  transformation devised to rewrite the  grand canonical 
partition function (GCPF) of the inhomogeneous fluid as the partition function  of a statistical 
field theory involving the stochastic real field $\phi$. The method was mainly used for 
homogeneous fluids in refs\ \cite{jmc_1,CV} and is considered here to study a planar interface. 
The theory is studied at the one-loop level where everything can be achieved analytically. 
The link between correlations of the field  and density correlations are established as well as 
the expressions for the surface tension and the surface structure factor. Technically, this is done 
{\it via} the determination of the eigenvalues and eigenfunctions of the second functional 
derivative of the mean field KSSHE Hamiltonian, which can be done exactly, as well known,
for the one-dimensional kink \cite{chaikin}. Taking into account the whole spectrum of 
eigenstates proves to be of the outmost importance to obtain the correct result. We thus obtain 
a satisfying picture of the liquid vapor interface, at least qualitatively, even at this simple 
level of description. 
One of the salient points of  our work is that we recovers both the CWT and its
first extension, namely the appearance of the bending rigidity factor $k$ and
the coupling between surface and bulk density correlations in $S(q)$, without invoking 
any {\it ad hoc} phenomenology. 
%
%%%% Section 1 apres cette ligne
\section {KSSHE transform and mean-field approximation}
\label{ksshe}
The important steps leading to the appropriate statistical field theory 
for a liquid are outlined in Appendix  \ref{appendix};  the 
interested reader is referred to ref.\cite{jmc_1,CV,siegert} for deeper details
and to ref. \cite{brill_bulk} for an alternative formulation.
We consider a simple fluid whose pair interaction potential includes a hard
sphere (HS) repulsive part and a soft attractive part, denoted by v(r). Let
\mbox{$w(r) = -\beta$v(r)} as usual and suppose that only purely attractive 
potentials are considered ($w$ is a positive definite operator, {\it i.e.} 
$\mt{w}(k)$ > 0).
We work in the grand canonical ensemble, namely at constant chemical potential
\mbox{$\nu = \beta\mu$}, and we consider the grand canonical partition function
(GCPF)  related to the grand potential \mbox{$\Xi = \exp(-\beta\Om)$}.
Following the notations of  appendix \ref{appendix} we obtain for $\Xi$ 
\equa{ksshe_1}
\Xi[\nu] &=& \frac{1}{N_w}\int \mathcal{D} \phi \exp(-H[\phi]) \nonumber \\
H &=& \frac{1}{2} \phi \cdot w^{-1} \cdot \phi  - \ln(\Xi_{\mathrm{HS}}[\bar{\nu}+\phi]) ,
\auqe
where $\phi$ is a real scalar field, $\mathcal{D} \phi$ a functional integration measure,
 $\bar{\nu}$ = $\nu - w(0)/2$ and, finally,  $\Xi_{\mathrm{HS}}[\nu]$ is the hard sphere grand partition 
functional of the local chemical potential  $\nu(\vec{r})$. A Landau-Ginzburg form 
is obtained from a functional Taylor expansion of $\ln(\Xi_{\mathrm{HS}})$ around a conveniently 
chosen  reference chemical potential of the hard sphere fluid, $\nu_0$
(see Appendix \ref{appendix}).
The propagator is expanded up to order $k^2$ in Fourier space and we are left with
\begin{subequations} 
\label{ksshe_2}
\equa{ksshe_2a}
H[\phi] = H_0 + \int
\left(\frac{K_2}{2} \left( \frac{\partial \phi}{\partial \vec{r}} \right)^2
+ \frac{K_0}{2} \phi^2 + V(\phi)
-  B \phi   \right) d\vec{r}  
\auqe
 \equa{ksshe_2b}
V(\Phi) = \sum_{n \geq 3} \frac{u_n}{n!}\phi^n
\auqe
\equa{champ_ext}
B(x) = \int dy \,w^{-1} (x,y)\gD \nu(y) + \rho_{\mathrm{HS}}\left[ \nu_0 \right](x) \makebox[1.5 cm]{with}
\gD\nu = \bar{\nu} - \nu_0.
\auqe
\end{subequations}
where the coupling constants $K_0, K_2$ and $u_n$ depend only on the
equation of state of the hard sphere fluid and the potential $w$ (see below). 
This formulation allows an exact mapping between the densities and their 
correlations on the one hand and the mean value and the correlations of the 
field on the other hand, as examplified in  Eq.\ \eqref{rho_phi} in the appendix.
For instance at coexistence, the densities $\rho_l$ and $\rho_g$ of
the two phases correspond to the two values $<\phi>_l$ and $<\phi>_g$.
The value of $\nu_0$ is chosen in such a way that $u_3(\nu_0)$
vanishes and the coexistence condition between the
liquid and vapor phases in the absence of external field is $B = 0$. 
$K_0$ is related to the deviation from the
critical temperature: $t = (T_c - T)/T_c~\propto~(-K_0)$. \\
In the inhomogeneous system,
we assume a special realization of the two phases coexistence:
we impose explicitly the occurrence of a bulk liquid and a bulk gas,
at densities $\rho_l$ and $\rho_g$ separated by a planar
surface located at \mbox{$z = z_0$}. The system is supposed to be bounded
in the direction $z$ according to $(z_0 - L) < z < (z_0 + L)$, 
$L$ being macroscopically large though formally finite.
The localization of the interface is obtained through the introduction of
a small valued external field $h_{ext}(z-z_0)$, odd in $(z-z_0)$.
Since the mean field profile takes constant values in both bulk phases, so is the case for
$h_{ext}$ which therefore plays the same role as  the truncated gravitational
field used in \cite{romero_92}. The advantage of our choice will appear below;
we emphasize that we do not aim to study gravitational effects but rather to fix the 
location of the interface at \mbox{$z =z_0$}. The inhomogeneous mean field theory 
or saddle point equation, \mbox{$\de H[\phi]/\de\phi = h_{ext}(z-z_0)$}, is solved 
by exploiting the fact that we look for a monotonous solution for the mean field 
profile $\phi_c(z)$ = $\phi_{\mathrm{MF}}(z)/\phi_{b}$, where $\phi_{b}$ is the value of 
$\phi_{\mathrm{MF}}(z)$ when $z \tend z_0 + L$, and by expanding $h_{ext}$ as a function 
of $\phi_c(z)$, namely $h_{ext}(z) = h_{ext}(\phi_c(z))$. The mean field equation 
then reads
\equa{mf1}
K_2\ddf = K_0 \phi_{\mathrm{MF}} + \frac{\partial V}{\partial\phi}(\phi_{\mathrm{MF}})
 - h_{ext}(\phi_{c})
\auqe
Since we consider a local approximation for the \mbox{$n \geq 3$}  HS kernels
(see Eq. (\ref{g_hs_nt})),  the r.h.s of Eq.(\ref{mf1}) is a polynomial in 
\mbox{$\phi_{\mathrm{MF}}(z-z_0)$}. Limiting $V(\phi)$ to the first non vanishing term, 
namely $V(\phi)$ = $(u_4/4!)\phi^4$, the analytical form for $\phi_{\mathrm{MF}}$ is not
modified by the introduction of an external field if the later expands as 
$h(\phi_c) = h_1 \phi_c + (h_3/3!) \phi_c^3$ ($h_1$, $h_3$ constants),
 as already noted by Zittartz \cite{zitt} when  $h(\phi_c) = h_1 \phi_c$ 
(This results holds in a more general way for a potential $V(\phi)$ even in $\phi$ and 
an external field odd in $\phi_c$ where the highest power of $V(\phi)$ is that of 
$h(\phi_c)$ plus one). The solution of the mean field equation in presence of the external field,
$\phi_{\mathrm{MF}}(K_2,K_0,u_4,h_{ext})$, then coincides with the solution in the
absence of external field, with modified values of the coupling constants
$K_0$ and $u_4$.
\equa{mf2}
\phi_{\mathrm{MF}}^{(K_2,K_0,u_4;h_{ext})}(z) &=& \phi_{\mathrm{MF}}^{(K_2,K'_0,u'_4;0)}(z)
\mbox{\; ; } \blanc K'_0 = K_0 - \delta K_0 \blanc u'_4 = u_4 - \delta u_4 \nonumber \\
\mbox{with \;}     \delta K_0 &=& h_1 \left( \frac{u'_4}{-6K'_0} \right)^{1/2}  
           \blanc  \mbox{and \;}  \delta u_4 = h_3 \left( \frac{u'_4}{-6K'_0} \right)^{3/2}  
\auqe
The solution of equation (\ref{mf1}) is 
$\phi_{\mathrm{MF}}(z) = \phi_b\phi_c((z-z_0)/l)$ with $\phi_c(x)$ = tanh(x), 
$\phi_b = (-6K'_0/u'_4)^{1/2}$ and $l = (2K_2/-K'_0)^{1/2}$ is the bulk correlation 
length which coincides also with the intrinsic interface width. We emphasize 
that we deal only with subcritical temperatures where \mbox{$K'_0 < 0$}.
We first have to check the mapping between the physical external field seen 
by the liquid, say $\Psi(z)=\be V_{ext}(z)$, and the external field which enters 
the effective Hamiltonian of equation (\ref{ksshe_2a}). We thus have to solve
\equa{champ_ext3}
B(1) = B_0 -\int w^{-1}(1,2) \Psi_{ext}(2)d2 \equiv h_{ext}(1)
\auqe
$B_0$ given by Eq.(\ref{champ_ext}) is the constant external field appearing 
in the effective Hamiltonian when $h_{ext}$ vanishes. The coexistence condition 
at constant chemical potential still reads $B_0 = 0$. Since 
the  propagator has been expanded up to  order $k^2$ (see appendix \ref{appendix}),
$w^{-1}$ is given in real space, at  the same order of approximation, by 
\equa{w-1}
w^{-1} (1,2) =
\frac{1}{\mt{w}_0}\left[ 1 + \frac{\mt{w}_2}{\mt{w}_0}\Delta_1 \right] \delta(1,2)
\auqe
where $\mt{w}_0$ and $\mt{w}_2$ denote the coefficients of the expansion at order
$k^2$ of the Fourier transform, $\mt{w}(k)$, of $w$.
By limiting ourselves  to functions depending only upon $z$, we thus have 
\equa{champ_ext4}
\int w^{-1}(1,2) \Psi_{ext}(2)d2 = 
\frac{1}{\mt{w}_0} \left[ 1 + \frac{\mt{w}_2}{\mt{w}_0} \frac{\partial^2}{\partial z_1 ^2} \right] \Psi_{ext}(1) 
& = & h_{ext}(z_1)
\auqe 
We introduce the bulk correlation length, $l$ and
we look for a solution depending on $t$ = $\phi_c(z/l)$ for both $h_{ext}(z_1)$ and 
$\Psi_{ext}(z_1)$. Equation (\ref{champ_ext4}) then reads
\equa{champ_ext5}
\left( 
(1-t^2)^2 \frac{\partial^2}{\partial t^2} - 2t(1-t^2)\frac{\partial}{\partial t}
+ l^2 \frac{\mt{w}_0}{\mt{w}_2} 
\right) \Psi = -\frac{\mt{w}_0^2}{\mt{w}_2}l^2 h_{ext}
\auqe
A one to one mapping with $h_{ext}$ limited to the third power in $\phi_c$ 
is obtained when \mbox{$\partial ^2\Psi/\partial t^2 = 0$} and we get
\equa{champ_ext6}
\Psi(z) = \Psi_1\phi_c(z) \blanc; \blanc
h_{ext}(z) &=& h_1\phi_c(z) + \frac{h_3}{3!}\phi_c(z)^3
\nonumber \\
h_1 &=& -\Psi_1\frac{1}{\mt{w}_0}\left(1 - \frac{2}{l^2}\frac{\mt{w}_2}{\mt{w}_0} \right)
\nonumber \\
h_3 &=& -\Psi_1 \frac{2}{l^2}\frac{\mt{w}_2}{\mt{w}_0}
\auqe
Formally $l$ is finite but clearly  the present approach should be valid only but in  the vicinity of the 
critical point where \mbox{$l \tend \infty$} and we   can therefore neglect the term $h_3$. \\
The mean field surface tension follows from the identity 
$\Omega_{\mathrm{MF}}[\nu]$ = $H[\phi_{\mathrm{MF}}(z)]$ which is easily calculated
by using both the equation satisfied by $\phi_{\mathrm{MF}}$ and the expression of the
external field contribution which takes a simple form with the particular
choice for $h_{ext}$, with the result
\equa{ga_mf}
\be\s^2\ga_{MF} = 4\s^2(-2K_2 K_0^{'3}/u_4^{'2})^{1/2}
+ 2\s^2\left(\frac{3K_2}{u'_4}\right)^{1/2}h_1\left(1 + \frac{h_3}{3h_1}\right)
\nonumber \auqe
which is similar to the expression given by Brilliantov \cite{brill_surf}
in the absence of the external field and differs from that given by Zittartz \cite{zitt} 
because of a different choice for the external field as a function of $\phi_c$.
When $h_3 = 0$ and $h_1 \tend 0$ at finite $K_0$ we get
\equa{ga_mf_lim}
\be\s^2\ga_{MF}(K_0, K_2, h_1 \tend 0)
\tend \be\s^2\ga_{MF}(K_0, K_2, h_1 =  0) +
\frac{\s^2 4(K_2)^{1/2}\phi_b}{(-3K_0)^{1/2}}h_1
\auqe
For $t~\tend~0$, one recovers the mean field exponent,  \mbox{$\ga_{MF} \sim t^{3/2}$}. \\
%%%%  Section 2 apres cette ligne
\section {One loop equations}
\label{one_loop}
In order to go beyond the mean-field  approximation, we expand the Hamiltonian 
about  $\phi_{\mathrm{MF}}(z)$ and consider the Gaussian approximation. 
From here, we drop the cubic term in the external field (\textit{i.e.} we set $h_3 = 0$) 
and, in order to  unclutter notations, we define
\mbox{$\hb = h_1/\phi_b$}. Without any lost in generality, we also choose \mbox{$z_0 = 0$}.
 The first correction stems from the second order term and we have ($z_0 = 0$),
\equa{gauss_1}
H[\phi = \phi_{\mathrm{MF}} + \chi] &\simeq&
H[\phi_{\mathrm{MF}}] + \frac{1}{2}\int \chi(1) H^{(2)}(1,2) \chi(2) d1 d2
\auqe
The operator $H^{(2)}$ = $\delta^2H/\delta\phi(1)\delta\phi(2)$
is diagonalized, after a Fourier transform parallel to the surface,
in the set of the eigenfunctions ${\varphi_{\la}}$ solution of the eigenvalues
equation ($\ep_{\la} = K_2(q^2 + \hb + \om_{\la}/l^2)$)
\equa{vp_1}
K_2 \left ( -\frac{\partial^2}{\partial z^{*2}} +
6(\tanh^2 (z^*) - 1) + (4 - \om_{\la})
\right) \varphi_{\la}(z^*)
= 0 \makebox[1.5 cm]{} z^* = z/l
\auqe
the solutions of which are known \cite{zitt,brez,infeld}. We emphasize that keeping only
the linear term in the external field leads to the same Schrodinger - like equation
(\ref{vp_1}) as in the absence of external field and the only change is a shift of
the eigenvalues $\ep_{\la}$ of $\hb$. The spectrum of
eigenfunctions includes two bound states, $\varphi_0~=~C_0/\cosh^2(z/l)$ and
$\varphi_1~=~C_1\sinh(z/l)/\cosh^2(z/l)$, with $\om_0$ = 0, $\om_1$ = 3 respectively
and a subset of unbounded states, or continuum
spectrum $\fik$, with $kl$ = \mbox{$\sqrt{\om_k-4}$} which behave as plane waves
in the bulk phases, {\it i.e.} far from the interface, and are given by
\equa{phi_k}
\fik =
C(k)e^{ikz} \left[ 2 - k^2l^2 - 3 i~kl\tanh(z/l) + 3~(\tanh^2(z/l) - 1) \right]
\auqe
Notice that the shift of the eigenvalues due to the external field is of crucial
importance especially for the lowest bounded state $\varphi_0$ the energy of which 
$\epsilon_0$ tends to a finite value, $K_2\hb$ when \mbox{$q \tend 0$}.
 As we shall see in the sequel this corresponds to a pinning of the interface by the field. 
It is important to note that 
\mbox{$\varphi_0(z) \propto \partial\phi_{\mathrm{MF}}(z)/\partial{z}$}.
The constants $C_0$, $C_1$ and $C(k)$ are determined in order to normalize
the $\varphi_{\la}$ on the interval $[-L, L]$.
The whole spectrum of eigenstates must then be orthogonalized. For the bound states
this is a direct consequence of the eigenvalues equation (\ref{vp_1}) in the limit 
\mbox{$L/l\tend\infty$}, {\it i.e.} disregarding terms or order $\exp(-2L/l)$. On the other 
hand the orthogonalization of the subset of unbounded eigenstates is no more a 
consequence  Eq. (\ref{vp_1}) but  rather follows from the boundary condition
\equa{norma_1}
\left[ \varphi_k\varphi^{\star}_{k'} \right]_{-L}^{L} = 0 \blanc \mbox{for}\blanc k \ne k'
\auqe
from which we get the non trivial dispersion relation leading to the density of states ($x = k~l$)
\equa{dos}
n(k) = \frac{L}{\pi} - \frac{l}{\pi} \left[
\frac{1}{(1+x^2)} + \frac{2}{(4+x^2)} \right] =
\frac{L}{\pi} + l~f(x)
\auqe
The closure relation then  follows
\equa{clos}
\int_{-\infty}^{+\infty} \frac{dk}{2\pi} \left( n(k) \abs{\varphi_k(z)}^2 -1 \right) 
+ \varphi_0^2(z) + \varphi_1^2(z) =0 
\auqe
It is important
to note that \mbox{$\int_{-\infty}^{\infty} f(x) dx$ = -2} which proved useful
in all calculations. Zittartz \cite{zitt} have already got this density of
states but without explicitly mentioning the need of the orthogonalization.
In \cite{dicap} a similar kind of dispersion relation was obtained in the
modeling of the charge density profile of electrolytes in the framework of
another field theory. We emphasize that the orthogonalization of the eigenstates
is of crucial importance for the calculation of gaussian functional integrals.
In order to calculate $\ln(\Xi)$ and the correlation functions, we write
the fields $\chi(\vec{r})$ in the basis which diagonalizes the operator
$H^{(2)}$, namely a Fourier transform in the $\vec{s}$ - direction  and a
projection on the $\fix{\lambda}$ where $\{\lambda\}$ = $\{n=0,1; k\}$ denotes
the whole spectrum of eigenstates. From usual gaussian functional integrals
\cite{binney} we get
\equa{log_xi}
\ln(\Xi) = - H[\phi_{\mathrm{MF}}] +
\frac{V}{2} \int \frac{d_3\vec{k}}{(2\pi)^3}
~\ln\left(\frac{\hat{w}^{-1}(k)}{\ep_k(q)}\right)
\\ -
\frac{S}{2} \int \frac{d\vec{q}}{(2\pi)^2} \left(
\sum_{n}~\ln \left( \ep_{n}(q) \right)
+ \int dk f (kl)~\ln \left( \ep_{k}(q) \right)
\right)
\nonumber \auqe
where the $\ep_{k}$ coincide with the eigenvalues of the second functional derivative 
of the Hamiltonian for the $\phi^4$ homogeneous model taken at 
$\phi~=~\phi_{\mathrm{MF}}(\pm~L)$. Therefore, the volume term of $\ln(\Xi)$ in 
(\ref{log_xi}) is nothing but \mbox{$\be pV$} at the one loop approximation \cite{jmc_1}.
We finally write the surface term of
$\ln(\Xi)$ in term of the surface tension $\ga=[\ln(\Xi_b) - \ln(\Xi)]/S$ =
$\ga_{MF} + \ga^{(1)}$. We obtain ($\ga^{(1*)}$ = $\be\ga^{(1)}\s^2$)
\equa{gamma_1}
\ga^{(1*)} = \frac{1}{8\pi} \left[
(t-1) \ln(1 - \frac{2}{\sqrt{t}+1}) + (t-4)\ln(1 - \frac{4}{\sqrt{t}+2})
-6\sqrt{t}
\right]_{a+(q_ml)^2}^{a+(q_Ml)^2}
\nonumber \\
\auqe
where $q_m$ and $q_M$ are the lower and upper bound respectively of the integral
over $q$ and $a$ = $(4+\hb)$. A similar result was obtained in \cite{zitt} for a spin 
model. This result differs from that obtained in \cite{lipo_5} where only the $n = 0$ 
eigenstate is kept. In the limit \mbox{$\hb \tend 0$}, we get
\equa{gamma_1_lim}
\ga^{(1*)}(\hb) \tend \ga^{(1*)}(\hb = 0) -(1/8\pi) \hb\ln(\hb).
\auqe
We now consider the calculation of the field two-body correlation functions,
\linebreak
\mbox{$G_{\phi}(1,2) = <\chi(1)\chi(2)>_{H}$}.
More precisely, we focus on the Fourier transform parallel to the surface,
$G_{\phi}(z_1,z_2,q)$. For this we have to calculate the sum
\equa{corr1}
\sum_{\la}\frac{\varphi_{\la}(z_1) \varphi_{\la}^{*}(z_2)}{\ep_{\la}(q)} =
\sum_{n=0,1}\frac{\varphi_{n}(z_1) \varphi_{n}(z_2)}{\ep_{n}(q)}
+ \int\rho(k)\frac{\varphi_{k}(z_1) \varphi_{k}^{*}(z_2)}{\ep_{k}(q)} dk
\nonumber \auqe
A contribution of the integral over $k$ cancels exactly the direct contribution
of the two bound states, which shows, once again, that the approximation
consisting in keeping only the $\fix{0}$ eigenmode is not sufficient. The result
is ($z_{>,<} = sup,inf(z_1,z_2)$)
\pagebreak[0]
\equa{corr_1}
G_{\phi}(z_1,z_2,q) = \nopagebreak[4]
\frac{9l}{2K_2}
\frac{\exp(-\abs{z_{12}}(x^2+4)^{1/2}/l) }{ x^2(x^2 + 3)(x^2+4)^{1/2} }
\nonumber \\ \nopagebreak[4]
\textrm{x}\left[1+ x^2/3 + ( x^2+4)^{1/2}\tanh(z_>/l) + \tanh^2(z_>/l) \right]
\nonumber \\ \nopagebreak[4]
\textrm{x}\left[1+ x^2/3 - ( x^2+4)^{1/2}\tanh(z_</l) + \tanh^2(z_</l) \right]
\auqe
where $x^2 = (q^2 + \hb)l^2$. Notice that formally this expression coincides with that obtained  in the absence 
of the external field, the only difference being the definition of $x$ which takes a finite value, $\hb$ when \mbox{$q \tend 0$}.
%
%%%%  Section 3 apres cette ligne
\section { Results and discussion}
\label{result}
\subsection{Capillary behavior and surface structure factor}
We started from the expansion of the effective Hamiltonian on the basis of the
eigenstates $\fix{\la}$. The first eigenstate, $\fix{0}$ is proportional to the
derivative of the mean field result $\phi_c(z)$, the proportionality constant
being determined by the normalization. If we keep only this first eigenstate,
the expansion of $\chi$ reads $\chi(\vec{s},z)~=~\xi(\vec{s})\fix{0}$. Hence the
field takes the form $\phi(z) = \phi_c(z-z_{int}(\vec{s}))$ and only the linear
term in the expansion of $\phi$ with respect to $z_{int}$ . remains. This
corresponds indeed to the so-called rigidly displaced profile approximation, where
$z_{int} = -a\xi(\vec{s})$ represents the fluctuating location of the interface.
The corresponding contribution to $H$ is a functional of \mbox{$z_{int}(\vec{s})$} which
defines an effective surface Hamiltonian given by
\equa{cwt_1}
H_s^{(0)}[z_{int}(\vec{s})] = \frac{K_2}{2a^2} \int d\vec{s}
\left[ (\partial_{\vec{s}}(z_{int}(s))^2 + \hb z_{int}(s)^2 \right]
\auqe

where we have used $\ep_0 = K_2(q^2 + \hb)$ since $\om_0$ = 0. This Hamiltonian 
pertains to the capillary wave models \cite{boulter_rev,zia_84,romero_92,diehl_01}.
In general the capillary wave Hamiltonian is obtained either from 
phenomenological arguments \cite{boulter_rev} or by restricting the fluctuations beyond
the mean field level to those deduced from a rigidly displaced mean field profile \cite{zia_84}.
In \cite{romero_92} a formulation based upon the direct correlation function of the
liquid state theory is used. In \cite{segovia} a supplementary 
\mbox{$z_{int}(s)^4$} term  in the surface Hamiltonian (\ref{cwt_1}) was introduced in order to 
account for curvature effects of the interface; this additional  term must vanish with the external field, 
\cite{diehl_01} and in any case cannot be generated in  the framework a 
one loop scheme. From the normalization of $\varphi_0$ and the analytic form of 
the mean field profile we deduce $K_2/a^2$ =$\s^2\be\ga_{MF} - \s^2l\phi_b^2\hb$
and accordingly we rewrite (\ref{cwt_1}) in the form
\equa{cwt_2}
H_s^{(0)} = \frac{1}{2}
\left[ \be\ga_{MF} - \s^2l\phi_b^2\hb \right]
\int d\vec{s}
\left[ (\partial_{\vec{s}}(z_{int}(\vec{s})))^2 +\hb z_{int}(\vec{s})^2 \right]
\auqe
which coincides exactly with the usual effective surface Hamiltonian
$H_{\mathrm{CWT}}[z_{int}(\vec{s})]$ of the CWT theory for the free surface,
when the external field vanishes.
Therefore, in the limit \mbox{$h_1 \tend 0$} we obtain the CWT as the lowest
approximation beyond the mean field approximation without invoking
phenomenological arguments.
The structure is characterized by the height-height correlation function,
$<z_{int}(\vec{s}_1)z_{int}(\vec{s}_2)>$ or its Fourier transform parallel to
the surface which defines the surface structure factor, $S_{\gD{z}\gD{z}}(q)$,
where $z_{int}(\vec{s})$ is the location of the interface relative to its mean
value. We consider $\int\chi(\vec{s},z) dz/\gD\phi_b$, where
$\gD\phi_b = \phi_{\mathrm{MF}}(L)-\phi_{\mathrm{MF}}(-L)$, as a measure of the
instantaneous location of the surface at $\vec{s}$, which amounts to define
the location of the surface from a constraint on the integral of $\chi$,
as is done in \cite{blokhuis} for the density profile. We are
then led to identify $\szz(q)$ = \mbox{$(\gD\phi_b)^{-2}\int G_\phi(z_1,z_2,q)dz_1dz_2$},
which corresponds to the $S_{ic}$ used in \cite{blokhuis}.
It is important to notice that the coupling with the bulk fluctuations are included
in the present formulation through the eigenstates of the continuum.
We also introduce the effective surface width, or surface corrugation,
\mbox{$\s^{eff} = \sqrt{<z_{int}(\vec{s}_1)z_{int}(\vec{s}_1)>}$}.
Henceforth  we consider only small values of $\hb$ and more precisely 
we relate the value of $\hb$ to an effective lateral size of system, according to 
\mbox{$\hb \sim L_x^{-2}$} as it will be justified from the analysis of the capillary 
wave limit. The behavior of $\szz(q)$ is analyzed from the function
\mbox{$\mt{g}(q) = \int dz_1\int dz_2 G_\phi(z_1,z_2,q)$}.
From (\ref{corr_1}) we get for the leading term of $G_\phi(z_1,z_2,q)$ when \mbox{$q\tend 0$}
\equa{g_q_zero_1}
G_{\phi}(z_1,z_2,q \tend 0) ~\simeq~ \frac{1}{K_2 (q^2 + \hb)} \varphi_0(z_1) \varphi_0(z_2)
+\frac{3l}{2K_2}  \frac{\hb}{q^2 + \hb} e^{-2\abs{z_{12}}/l}  \left[  f(z_1,z_2) \right]
\nonumber \\
\auqe
The first term of the r.h.s of Eq(\ref{g_q_zero_1}) corresponds to the $CWT$ limit 
especially when $\hb = 0$, and the correction, which vanishes at $\hb = 0$, plays a 
role only far from the interface where $\varphi_0(z_1) \varphi_0(z_2)$ is negligible 
and moreover when the two particles are on the same side of the interface, {\it i.e.} 
when $e^{-2\abs{z_{12}}/l}$ is not negligible. Thus we can rewrite the  \mbox{$q\tend 0$}
behavior of $G_{\phi}(z_1,z_2,q)$ as
\equa{g_q_zero_2}
G_{\phi}(z_1,z_2,q \tend 0) ~\simeq~ \frac{1}{K_2 (q^2 + \hb)} \varphi_0(z_1) \varphi_0(z_2)
+\frac{l}{2K_2}~f~e^{-2\abs{z_{12}}/l}
\auqe
where $f$ takes a constant value, $f = 1$, when both $\abs{z_1}, \abs{z_2} >> l$
and are of the same sign. The second term of the r.h.s of (\ref{g_q_zero_2})
is nothing but the limit at \mbox{$q\tend0$} of the bulk correlation function and is not to be
included in the capillary limit of the model. In any case, for sufficiently small values of $\hb$,
the bulk contribution to Eq. (\ref{g_q_zero_2}) becomes negligible at \mbox{$q = 0$}.  This
is made more precise below concerning the behavior of $\szz$. 
Let us denote by $G_{CW}$ the first term of the r.h.s. of Eq.(\ref{g_q_zero_2})
and by $\mt{g}_{CW}(q)$ its integral over $z_1$ and $z_2$.
It is easy to show from the relation already used between the normalization of
$\fix{0}$ and $\ga_{MF}$ that
\equa{s_q_zero}
\frac{1}{(\gD\phi_b)^2} \mt{g}_{CW}(q \tend 0)
\tend \frac{1}{(\ga_{MF} - \s^2 l\phi_b^2 \hb)(q^2 + \hb)}
\auqe
In the limit $\hb = 0$ this is exactly the CWT behavior, leading to the well known
logarithmic divergence of the squared surface corrugation for which one gets
$(\s^{eff})^2$ = \mbox{$(4\pi\ga_{MF})^{-1}\ln(q_{M}^2/q_{m}^2)$}.
The effect of the external field is to make the integral over $q$ entering in
the determination of $\s^{eff}$ finite at \mbox{$q \tend 0$} with the result
\equa{sig_eff}
(\s^{eff})^2  =
\frac{1}{4\pi\be\s^2\ga_{MF}}\ln \left(\frac{(2\pi)^2}{\s^2\hb} \right)[1 + O(\s^2\hb)]
\auqe
where we have assimilated $q_M$ to the inverse of the molecular size, $2\pi/\s$.
Since the result for the CW at the free interface in the absence of external field 
is a logarithmic divergence with the lateral size, \mbox{$\sim \ln(\s/L_x)$} \cite{bls},
this shows that, as we already mentioned, the effect of the external field,
is to pin the interface in such a way that the capillary wave are restricted
to a lateral scale $L_x \sim \hb^{-(1/2)}$.
It can be shown from (\ref{corr_1}) that the contribution to $\mt{g}(q)$ diverging
with the system size, due to the bulk correlations,
is exactly $(2L)$ times the integral of $G_b(z_{12},q)$,the correlation function
of the bulk phase, over $z_{12}$ If we keep only these two terms we get
\equa{corr_3}
\szz(q) \simeq
\frac{2L}{\gD\Phi^2}\int G_b(q,z_{12})dz_{12} + 
\frac{1 + (\s^2l\phi_b^2\ga_{MF})\hb}{\ga_{MF}(q^2+\hb)}
+ O(\hb^2)
\auqe
Given that the bulk term leads to a constant when $q \tend 0$,
we see that (\ref{corr_3}) presents a cross-over like behavior in terms of
wave vector $q$, where a threshold value $q_s$ naturally appears,
separating the capillary wave behavior at small values of $q$ from the
bulk like behavior at $q > q_s$, with $q_s$ given by
\equa{qs}
(q_s^2 + \hb)
~\sim~
\frac{\gD\Phi^2}{2L~\be\ga_{MF}K_2\int G_b(q,z_{12})dz_{12}}
\auqe
We see that this cross-over behavior holds only if $\hb$ is negligible compared to the
r.h.s.of Eq.(\ref{qs}) or in other words
\equa{qs_2}
\hb~l^2 < \frac{l}{2L}\frac{16\phi_b^2lK_2}{\be\ga_{MF}l^2}
\nonumber \auqe
and confirms the above mentioned separation of the capillary behavior from the bulk
correlations when this condition is fulfilled. We emphasize that $\hb^{1/2}$ can be
understood as the inverse of an effective lateral size pinning the interface.
The cross-over like behavior resulting from (\ref{qs}) is in agreement with that of
Refs. \cite{blokhuis,sim_binder,sedlmeier_09};
however, in the present formulation the bulk fluctuations come out of the
calculation through the continuum subset of eigenstates and have not to be added
to the field profile. Furthermore, from (\ref{corr_3}) we can drop
exactly the bulk contribution, and doing this we define a purely
interfacial contribution to $\szz(q)$ (see fig.(\ref{correl_int_bulk})).
\equa{s_int_1}
S_{int}(q) =
\szz(q) - (2L/\gD\phi_b^2)\int G_b(q,z_{12})dz_{12}
\auqe

Then the departure of $S_{int}(q)$ from its $1/q^2$ behavior allows us to isolate
the deviations from the capillary wave like behavior of $\szz(q)$. At this point, since
we focus on the deviation form the capillary - like behavior, we can consider $\hb=0$.
We define $\be(q)$ = \mbox{$(q^2S_{int}(q))^{-1} - \ga_{MF}$}. $\be(q)/q^2$ is found
nearly constant and thus lead us to define a bending rigidity of the interface,
\equa{kappa}
\kappa = \lim_{q\tend 0} ((1/q^2)\be(q)) \simeq ((1/q^2)\be(q))
\auqe
which takes a positive value, as it should be for the stability of the interface
\cite{dft_md}, as is also found from the simulation results of Ref. \cite{sedlmeier_09}
after a separation between bulk density correlations and interface fluctuations.
In other words, the small $q$ behavior of $S_{int}(q)$ is
\equa{s_int_2}
S_{int}(q) \simeq 1/[\ga_{MF} q^2 + \kappa q^4]
\nonumber \auqe
Here we find $\kappa$ = \mbox{$\kappa^*(\ga_{MF}\s^2)/(\s c)^2$}, with the
reduced value $\kappa^*$ = 0.288.

\subsection{Surface tension}
In this section we focus on the behavior of the one loop contribution to the
surface tension $\gamma$ in the case \mbox{$\hb=0$}, since its dependence
 on $\hb$
then vanishes (see Eq.(\ref{gamma_1_lim})).
The contribution $\ga^{(1)}$, given by (\ref{gamma_1}), depends on the two
bounds $q_m$ and $q_M$ of the wave vector $q$ which are related to the
relevant parameters of the interface: on the one hand $lq_M$ = $2\pi l/\s$
where $l/\s$ is the intrinsic width of the interface in unit $\s$,
$lq_M$ $\in$ $[1,\infty[$. On the other hand, $lq_m$ = $(2\pi l /L_x$ where
$L_x$ is the system actual lateral size; hence $l q_m$ = $(l q_M)(\s/L_x)$.
Thus $l q_M$ appears as a natural parameter, with $lq_M~\in~[2\pi, \infty]$
and $1/(lq_M)~\propto~\sqrt{t}$. We can now re-write $\ga^{(1*)}$ is a more
convenient form ($\s/L_x~\in~]0,1]$)
\equa{gamma_1b}
\ga^{(1*)} = \frac{\pi}{2(lq_M)^2} \left[ \mt{\ga}(4+(lq_M)^2) -
         \mt{\ga}(4+(lq_M\frac{\s}{L_x})^2) \right]
\auqe
$(L_x/\s)$ is either the actual lateral system size or the scale at which $\ga$ is
measured, for instance in a numerical simulation (see ref. \cite{tarazona}),
but in any case does not depend on $t$. We can note that whatever
the values of $\s/L_x$ or $(lq_M)$, $\s^2\ga^{(1)}$ remains finite and
more precisely
\equa{gamma_1lim}
\ga^{(1*)}_{L_x \tend \infty} (lq_M \tend \infty) \sim -\frac{6\pi}{l q_M}
\auqe
which $\tend$ 0 when $t \tend$ 0 as $t^{1/2}$.
The results for $\s^2\ga^{(1*)}$ are displayed in fig. (\ref{gamma1_red}).
We interpret $\s^2\ga^{(1*)}(\s/L_x)$ as the $q-$dependent contribution to $\ga$
with $q~=~(2\pi\s/L_x)$.
This means that the flat interface corresponding to the mean field approximation
is obtained when no fluctuation at all are taken into account, namely for
\mbox{$q \sim q_M$}. This differs from what is done in Ref. \cite{dft_md} (see also ref.
\cite{tara_comm}) where the contribution to $\ga$ due to the surface fluctuations
vanishes at \mbox{$q \tend 0$}.
The small $q$ behavior of $\ga^{(1*)}$ is easily obtained from (\ref{gamma_1b})
and yields
\equa{gamma1_lim2}
\ga^{(1*)}_{q\tend 0}
\simeq \ga^{(1*)}(0) + \frac{1}{8\pi}[(q\s)^2[a + \ln(\s/l)] - 2(q\s)^2\ln(q\s)]
\makebox[0.1\textwidth]{with} a = 3.48491 \nonumber \\
\auqe
from which we can deduce a crossover value of $q$ given by $q_{0}\s$ =
$(\s/l)e^{a/2}$,
separating the $q^2$ behavior from the $(q\s)^2\ln(q\s)$ behavior obtained for
\mbox{$q < q_{0}$} and \mbox{$q > q_{0}$} respectively. It is important
to note first that we always get an increasing $\ga^{(1)}(q)$ and
secondly that since $q_{0}$ is proportional to $l^{-1}$,
we get a plateau (corresponding to the $q^2$ dependence) at small
values of $\s/L_x$ only when $lq_M$ takes small values, {\it i.e.} for the lowest
temperatures (see figure (\ref{gamma1_red}). This behavior is in qualitative
agreement with the simulation results of \cite{tarazona}.
The term proportional to $q^2$ in the variation of $\ga^{(1)}$
with $q$ can be interpreted as resulting from the energy necessary to bend
the interface, and should be related to $\kappa$ obtained from the behavior
of $S_{int}(q)$ (see eq. (\ref{kappa}). This is not {\it a priori} the case
since we do not expect a fully coherence in the framework of a loop
expansion between the energetic and structure quantities.
\subsection {Density profile and density correlation function}
Now we go back to the mapping between the averaged field, $\phi$, and the
density in the fluid, $\rho$ on the one hand and between the field and the
two body density correlation function $G_{\rho}^{2T}$ on the other hand.
Formally we have from Eq.(\ref{rho_phi}))
\equa{map_1}
\rho(1) &=& \rho_{\mathrm{HS}}[\nu_0](1) - h_{ext}(1) + \int w^{-1}(1,2)\phi_{MF}(2) d2
\nonumber \\
G_{\rho}^{(2 T)} (1,2) &=& \int \int w^{-1}(1,1') w^{-1}(2,2') G_{\phi}(1',2') d1' d2'
                   - w^{-1}(1,2)
\auqe
We recall that  $G_{\rho}^{(2 T)} (1,2)$ is $\rho(1)\rho(2)~h(1,2) + \rho(1)\delta(1,2)$
where $h(1,2)$ is the usual total correlation function of liquid state theory\cite{Hansen}.
We easily get
\equa{map_rho2}
\rho(1) = \rho_{\mathrm{HS}}(\nu_0) + \Delta\rho_1 \phi_c(z_1) + \Delta\rho_3 \phi_c(z_1)^3
\auqe
where
\equa{map_rho3}
 \Delta\rho_1&=& \phi_b \left[\frac{1}{\mt{w}_0} - \frac{\mt{w}_2}{\mt{w}_0^2~l^2} -\hb \right] \nonumber \\
 \Delta\rho_3 &=& \phi_b \frac{\mt{w}_2}{\mt{w}_0^2~l^2}
\auqe
If we limit the expansion to the term linear in $h_1$ we recover the usual
tanh - like profile $(\rho_l + \rho_g)/2 + \tanh(z - z_0)(\rho_l - \rho_g)/2$.
Concerning the two body correlation function, since we limit the effective
Hamiltonian to a $\phi^4$ model, the bulk field correlation function is the
same in the two phases and as a result of equation (\ref{map_1}) this is
also the case for the density correlation function $G_{\rho}^{(2 T)}(1,2)$.
This is of course a drawback of the simplicity of the present model for the
description of the liquid vapor interface. To overcome this drawback, we may
have to consider a true inhomogeneous reference fluid, which impose to chose two
different values for the hard sphere reference chemical potential, $\nu_0$.
This is beyond the scope of the present work.
\section{Conclusion}

We have shown in this article  that the one-loop approximation  of the  KSSHE statistical field theory 
of simple liquids gives a coherent and qualitatively exact description of a planar liquid/vapor interface.
The mathematical treatment of the kink is subtle and requires
to take into account the whole spectrum of eigenvalues and eigenstates 
of the Gaussian Hamiltonian and not, as often and incorrectly proposed in the literature, to
restrict oneself  to its fundamental state.
Moreover,without invoking any  phenomenological description of the interface, we were able to
recover the usual CWT surface Hamiltonian as an approximation of the one-loop calculation.
We also show that we can extract a rigidity bending factor which takes a positive
value, in agreement with the requirement for the stability of the interface with
respect to fluctuations. On the other hand, we clearly control  the limit of the model
by computing  the two-body density correlation functions which appear to coincide in 
the liquid and vapor bulk phases, a drawback due
to the identity of the two phases in the field $\phi$ representation.

Nevertheless, qualitatively, our results are in good agreement with numerical simulations. 
We stress that no parameter is involved in the theory which can be worked out for any pair potential.
A quantitative agreement is expected only for long range pair potentials which we plan to
investigate by numerical simulations in future work.
%
% \eject
%%%%%%%%%%%%%%%%%%%%%%%%%%%%%%%%%%%%%%%%%%%
\appendix
\setcounter{equation}{0}
\renewcommand{\theequation}{A-\arabic{equation}}
\section{KSSHE Transform}
\label{appendix}
In this appendix we review the so-called Kac-Siegert-Stratonovich-Hubbard-Edwards (KSSHE)
\cite{Kac,siegert,Strato,Hubbard,Edwards}  transformation devised to rewrite the GCPF of a classical
fluid as the partition function  of a bosonic  statistical field theory;  more details are given in
refs~\cite{jmc_1,CV}.

We consider  the case of a simple three dimensional  fluid made of
identical hard spheres of diameter $\sigma$ with additional isotropic
pair interactions
$v(r_{ij})$  (\mbox{$r_{ij}=\Arrowvert x_i - x_j \Arrowvert$}, $x_i$ position of particle "$i$"). Since $v(r)$ is arbitrary in the core,
 i.e. for $r \leq \sigma$, we assume that $v(r)$ has
been regularized in such a way that  its Fourier transform
$\widetilde{v}(q)$ is a well behaved function of $q$ and that $v(0)$ is a finite quantity.
We denote by $\Omega \subset \R^{3}$ the domain of volume $V$
occupied by the molecules of the fluid.
The fluid is at equilibrium in the GC ensemble  and we denote
by $\beta=1/kT$ the inverse temperature ($k$ Boltzmann constant) and  $\mu$ the chemical potential.
For the sake of generality the particles are subject to an external potential $\psi(x)$ and
$\nu(x)=\beta (\mu-\psi(x))$ is the dimensionless
local chemical potential. We stick to notations usually adopted in standard textbooks on the theory
of liquids (see e.g. ref.~(\cite{Hansen})) and denote
 by $w(r)=-\beta v(r)$ the negative of the dimensionless pair interaction. Moreover we restrict ourselves
to  attractive interactions, \textit{i.e.} such that their Fourier transform
$\widetilde{w}(q) > 0$ is definite positive for all $q$.

In a given configuration
$\mathcal{C}=(N;x_1 \ldots x_N)$ of the GC ensemble
the microscopic density of particles reads
\begin{equation}
\widehat{\rho}(x\vert \mathcal{C})=
\sum_{i=1}^{N} \delta^{(3)}(x-x_i) \; ,
\end{equation}
and the GC partition function $\Xi\left[ \nu \right] $ can thus be written as
\begin{equation} \label{csi}
\Xi\left[ \nu \right] =
\sum_{N=0}^{\infty} \frac{1}{N!}
\int_{\Omega}d1 \ldots dn \; \exp\left(
-\beta V_{\text{HS}}(\mathcal{C})
+\frac{1}{2}
\widehat{\rho}\cdot w \cdot \widehat{\rho}
+  \overline{\nu}\cdot \widehat{\rho}  \right) \; ,
\end{equation}
where \mbox{$\rho\ \cdot w \cdot \rho$} is a short hand notation for
\mbox{$\int \rho(1) w(1,2) \rho(2) d1 d2$},
$i \equiv x_i $ and $di\equiv d^{3}x_i$. For a given volume $V$, $\Xi\left[ \nu \right]$
 is a function of $\beta$ and a convex functional of the local
 chemical potential $\nu(x)$\cite{Goldenfeld,Cai-conv}.
 In eq.\ (\ref{csi}) $\exp(-\beta V_{\text{HS}}(\mathcal{C}))$ denotes the hard
sphere contribution to the Boltzmann factor in a configuration $\mathcal{C}$  and
$\overline{\nu}=\nu+\nu_S$ where
$\nu_S= - w(0)/2$ is $\beta$ times  the  self-energy of
particles. From our hypothesis on $w(r)$,
$\nu_S$ is a finite quantity which depends however on the regularization of
the potential in the core.
%%%%%%%%%%%%%%%%%%%%%%%%%%%%%%%%%%

We now   recognize the Gaussian integral \cite{Wegner,Zinn}
 \begin{align}
\exp\left( \frac{1 }{2}
\widehat{\rho} \cdot w \cdot     \widehat{\rho} \right) &=\dfrac{1}{\mathcal{N}_{w}} \int \mathcal{D} \varphi \;
\exp\left( - \dfrac{1}{2} \varphi \cdot w^{-1} \cdot   \varphi
+\widehat{\rho} \cdot \varphi
\right)  \; ,
 \end{align}
where  $\varphi$ is  a real scalar field, $ \mathcal{D} \varphi$ a functional measure,
$w^{-1}(1,2)$ is the inverse of $w(1,2)$ in the sense of operators  and the the
 normalization $\mathcal{N}_{w}$ reads as
 \begin{align}
 \mathcal{N}_{w}&=  \int \mathcal{D} \varphi \; \exp \left( - \dfrac{1}{2} \varphi \cdot w^{-1} \cdot   \varphi
\right)   \nonumber \\
&=  \exp \left(-\dfrac{V}{2} \int \dfrac{d^{3}k}{(2 \pi)^{3}} \ln \widetilde{w}(k)\right) \label{norma}\; ,
 \end{align}
 It follows then that the GCPF $\Xi\left[ \nu \right]$ can  be re-expressed as the  functional integral
 \begin{eqnarray}
\label{action-KSSHE}
\Xi\left[ \nu \right] &=&
\mathcal{N}_{w}^{-1} \int \mathcal{D} \varphi \;
\exp \left(  -H \left[ \varphi,\nu \right] \right)
\end{eqnarray}
where the KSSHE action (or effective Hamiltonian) is expressed as
\begin{eqnarray}
 H \left[ \varphi,\nu \right]&=& \frac{1}{2} \varphi \cdot w^{-1} \cdot \varphi
-\ln \left\lbrace \Xi_{\text{HS}}\left[ \overline{\nu} + \varphi\right]\right\rbrace  \; .
\end{eqnarray}
The action $H \left[ \varphi,\nu \right] $ is non-canonical in the sense that the coupling
between  the field $\varphi$
and the external source $\overline{\nu}$ is non-linear as in usual Landau-Ginzburg
actions \cite{Zinn}. In order  to get a canonical field theory we perform
the translation $\phi = \varphi + \Delta \nu$ with $\Delta \nu =\overline{\nu} - \nu_{0}$
where $ \nu_{0}$ is some arbitrary uniform reference chemical potential to be chosen
conveniently later. The Jacobian of the transformation is obviously equal to one and then,
performing  a Taylor functional expansion of the grand potential
$\ln\left\lbrace  \Xi_{\text{HS}}\left[ \nu_0 + \phi\right]\right\rbrace $ about $ \nu_{0}$
we obtain a standard scalar field theory \cite{jmc_1,CV}, characterized by the partition function
 \begin{eqnarray}
\label{action-cano}
\Xi^{\star}\left[ B \right] &=&
\mathcal{N}_{w}^{-1} \int \mathcal{D} \phi \;
% \exp \left(  \frac{1}{2} \phi \cdot \Delta^{-1} \cdot \phi -\mathcal{V}_{I}\left[ \phi \right]  + B \cdot \phi
\exp \left(  \frac{1}{2} \phi \cdot \Delta^{-1} \cdot \phi - V \left[ \phi \right]  + B \cdot \phi
 \right)  \; ,
\end{eqnarray}
where the propagator $\Delta(x,y)$ and the interaction $V[\phi]$ are given by
\begin{subequations}
 \begin{align}
 \Delta^{-1}(x,y) &= w^{-1}(x,y) + G^{(2) \; \textrm{T}}_{\textrm{HS}}(\nu_0 \vert x,y)\; , \\ \label{inter}
V \left[\phi\right]  & =- \sum_{n=3}^{\infty}\dfrac{1}{n!} \int dx_{1} \ldots dx_{n} \;
                                 G^{(n) \; \textrm{T}}_{\textrm{HS}}(\nu_0 \vert x_{1}, \ldots, x_{n}) \phi(x_{1} )\ldots \phi(x_{n} )\; ,
 \end{align}
\end{subequations}
where the $ G^{(n) \; \textrm{T}}_{\textrm{HS}}$ are the truncated n-body correlation functions of a HS fluid
at chemical potential $\nu_0$ \cite{Hansen}
\begin{equation}
 G^{(n) \; \textrm{T}}_{\textrm{HS}}(\nu_0 \vert x_{1}, \ldots, x_{n}) =
\dfrac{\delta^{ n } \ln \Xi_{\textrm{HS}}\left[ \nu_0 \right] }{\delta \nu_0(x_{1}) \ldots  \delta \nu_0(x_{n}) } \; ,
\end{equation}
which are supposed to be exactly known, and where, finally,   the external source or  ''magnetic'' field $B(x)$ of
the theory is related to the
local  chemical potential through the linear relation
\begin{equation}
\label{B}
 B(x) = \rho_{\textrm{HS}}(\nu_0) + w^{-1}(x,y)  \Delta \nu(y) \; ,
\end{equation}
where Einstein convention of summation over repeated indices -here continuous variable $y$- was adopted.
The free energies of the non-canonical and canonical theories differ essentially by a
quadratic form, \text{i.e.}
 \begin{align}
  \ln \Xi\left[ \nu \right] &=  \ln \Xi_{\textrm{HS}}\left[ \nu_0 \right] -\dfrac{1}{2}
\Delta \nu   \cdot w^{-1} \cdot \Delta \nu +   \ln \Xi^{\star}\left[ B\right] \: ,
 \end{align}
which allows to easily  relate the correlations of the density   to that of the field. With the definitions
\begin{subequations}
 \begin{align}
   G_{\rho}^{(n) \; \textrm{T}}(\nu \vert x_{1},\ldots, x_{n})&=
         \dfrac{\delta^{ n } \ln \Xi \left[ \nu \right] }{\delta \nu(x_{1}) \ldots  \delta \nu(x_{n}) }  \; , \\
  G_{\phi}^{(n) \; \textrm{T}}(B \vert x_{1},\ldots, x_{n})&=
 \dfrac{\delta^{ n } \ln \Xi^{\star}\left[ B \right] }{\delta B(x_{1}) \ldots  \delta B(x_{n}) }  \; ,
 \end{align}
\end{subequations}
 one finds \cite{jmc_1,CV}
\begin{subequations}
 \label{rho_phi}
 \begin{align}
  \rho(x) &= w^{-1}(x,y)  \left\lbrace <\phi>(y) - \Delta \nu(y) \right\rbrace \nonumber \\
             & = w^{-1}(x,y) <\varphi>(y) \; ,\\
 G_{\rho}^{(2) \; \textrm{T}}(x,y)&= - w^{-1}(x,y) + w^{-1}(x,x^{'})w^{-1}(y,y^{'}) G_{\phi}^{(2) \; \textrm{T}}(x^{'},y^{'}) \; ,\\
 G_{\rho}^{(n) \; \textrm{T}}(x_{1},\ldots, x_{n})&=  w^{-1}(x_{1},x_{1}^{'}) \ldots w^{-n}(x_{1},x_{n}^{'})  \times \nonumber \\
                         & \times G_{\phi}^{(n) \; \textrm{T}}(x^{'}_{1},\ldots,  x^{'}_{n}) \textrm{ for } n \geq 3 \; .
 \end{align}
\end{subequations}
Note that the \textit{truncated} n-body correlation functions of the fields $\phi$ and $\varphi$ coincide for
$n\geq 2$ since  the two fields differ by a simple additional function.

At this point we introduce some approximations. First we adopt the ``point of view of Sirius''
and consider the kernels  $G^{(n) \; \textrm{T}}_{\textrm{HS}}(\nu_0 \vert x_{1}, \ldots, x_{n}) $
to be  short-range functions of their arguments when compared  with  the range of variations of field or density
correlations near the critical point. More precisely we thus assume that, for $n \geq 3$, one has
\begin{equation} \label{g_hs_nt}
 G^{(n) \; \textrm{T}}_{\textrm{HS}}(\nu_0 \vert x_{1}, \ldots, x_{n}) \approx \beta P_{\textrm{HS}}^{(n)}(\nu_0) \delta (x_n,x_1) \ldots
\delta (x_2,x_1) \; ,
\end{equation}
where  $P_{\textrm{\textrm{HS}}}^{(n)}(\nu_0)$ denotes the n-th derivative of the HS
pressure with respect to the chemical potential $\nu_0$. Moreover, neglecting
high-order contributions to $V\left[\phi\right]$ we get the somehow sketchy interaction
\begin{equation}
\label{V}
V\left[\phi\right] \approx \dfrac{u_3}{3!} \int dx \;  \phi^{3}(x) + \dfrac{u_4}{4!} \int dx \; \phi^{4}(x) \; ,
\end{equation}
where $u_n = - \beta P_{\textrm{HS}}^{(n)}(\nu_0) = - \rho_{\textrm{HS}}^{(n-1)}(\nu_0) $,
which should be valid but very close to the critical point.
Note that one should have $u_4 >0 $ for the action to be bounded from below at large
fields so that the theory is well-behaved.

A similar approximation is devised for the propagator for which we adopt a gradient expansion up to second order,
 more conveniently written in Fourier space as
\begin{subequations}
 \begin{align}
  \widetilde{\Delta}^{-1}\left( q_1, q_2 \right) &= (2 \pi)^{3} \delta^{(3)}\left( q_1+q_2\right)  \widetilde{\Delta}^{-1}\left(q_1 \right) \; , \\
  \widetilde{\Delta}^{-1}(q)&= \widetilde{w}^{-1}(q) +\widetilde{ \mathcal{C}}^{(2)}_{\textrm{HS}}(q) \nonumber \\
                                            &= K_0 + K_2 q^2 + \mathcal{O}(q^4) \; ,
 \end{align}
\end{subequations}
where the Ornstein-Zernicke relation $ \widetilde{ \mathcal{C}}^{(2)}_{\textrm{HS}}(q) =-1/ \widetilde{G}^{(2)\; \textrm{T}}_{\textrm{HS}}(q)$
defines  the Fourier transform of the HS two-body direct
correlation function (the definition of which includes an  ideal gas contributions so that $\widetilde{\mathcal{C}}^{(2)}_{\textrm{HS}}(q) =
\widetilde{c}_{\textrm{HS}}(q) - 1/\rho$,
$\widetilde{c}_{\textrm{HS}}(q)$  usual direct correlation function \cite{Hansen}.) A short calculation gives
\begin{subequations}
\label{K0K2}
 \begin{align}
  K_0 &= \dfrac{1}{\widetilde{w}(0)} - \rho_{\textrm{HS}}^{(1)}(\nu_0) \; , \\
  K_2& = \dfrac{1}{2}\left\lbrace
- \dfrac{\widetilde{w}_{2}}{\widetilde{w}(0)^{2}}
   +     \widetilde{c}_{ \textrm{HS} \;,2} \;   \rho_{\textrm{HS}}^{(1)}(\nu_0)^{2}     \right\rbrace  \; ,
 \end{align}
\end{subequations}
where   $\widetilde{w}_{2}$ and  $ \widetilde{c}_{ \textrm{HS} \;,2} $ are the second derivatives
of  $\widetilde{w}(q)$ and $ \widetilde{c}_{ \textrm{HS}}(q) $
 with respect to $q$ at $q=0$ respectively.
The quantities entering the coupling constants $K_0$ and $K_2$ in eqs~\eqref{K0K2} requires
the knowledge of the equation of state and pair correlation functions of the HS fluid, which can be
done in the framework of  Percus-Yevick theory for instance \cite{Hansen}.

At this point we choose $\nu_0$ such that $u_3= - \rho_{\textrm{HS}}^{''}(\nu_0)=0$ so that
the  action $\mathcal{S}$ of field $\phi$ reduces exactly
 to that of a  $\phi^4$ Landau-Ginzburg model, \textit{i.e.}
\begin{align}
 \mathcal{S}&=-\phi \cdot B + \int dx \; \left\lbrace
\dfrac{1}{2} K_0 \left(\nabla \phi \right)^{2}
+ \dfrac{1}{2} K_2 \phi^{2}
+ \dfrac{u_4}{4!}  \phi^4 \right\rbrace  \; .
\end{align}
A last remark is in order. As discussed in ref.~\cite{jmc_1} the chemical potential
$\nu_0\approx-0.025$ is uniquely defined (a consequence of the expected -and satisfied- convexity of 
the function $\nu \longmapsto \rho^{'}_{\textrm{HS}}(\nu)$)
and such that $\rho_{\textrm{HS}}(\nu_0)\approx0.25$, $\rho_{\textrm{HS}}^{(1)}(\nu_0)\approx 0.09$
and $u_4>0$.

%
%\newpage
%%%%%                        Biblio
\begin{singlespace}

\end{singlespace}
\eject
\renewcommand{\thefigure}{\arabic{figure}}
% \section*{Figure captions}
%\begin{itemize}
%\item [Figure \ref{correl_int_bulk}]
%Log-log plot of $\mt{g}^*(x)$ versus $x~=~ql$ for $\hb = 0$ and
%$L/l$ = 100, 80, 60, 40 and 20 from top; interfacial contribution
%$\gD\phi_b^2S_{int}^*(x)$, bottom; CWT limit, straight line.
%
%\item[Figure \ref{gamma1_red}]
%$\gamma^{(1)}(\sigma/L_x)/\abs{\gamma^{(1)}(0)}$ in terms of $\sigma/L_x$
%for \mbox{$lq_M/(2\pi)$ = 1}; 2; 5 and 50 from bottom to top.
%\end{itemize}
%
\eject
\begin{figure}[H]
\begin{center}
\includegraphics[width = 0.6\textwidth]{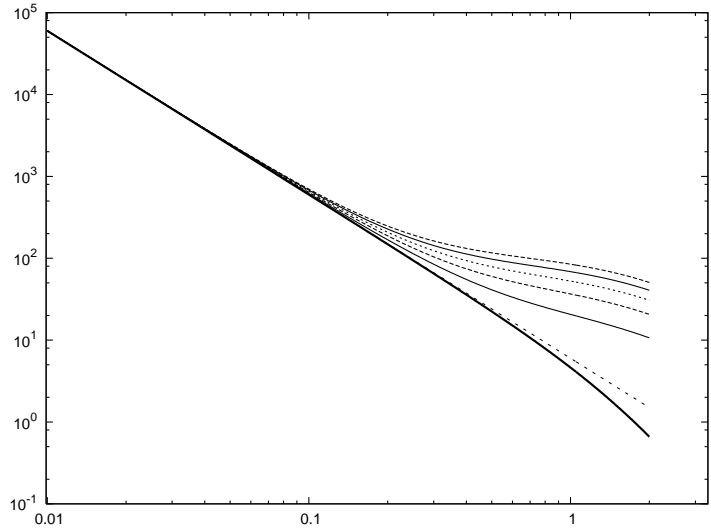}
\caption {\label{correl_int_bulk}
Log-log plot of $\mt{g}^*(x)$ versus $x~=~ql$ for $\hb = 0$ and
$L/l$ = 100, 80, 60, 40 and 20 from top; interfacial contribution
$\gD\phi_b^2S_{int}^*(x)$, bottom; CWT limit, straight line.
}
\end{center}
\end{figure}
\begin{figure}[H]
\begin{center}
\includegraphics [width = 0.6\textwidth]{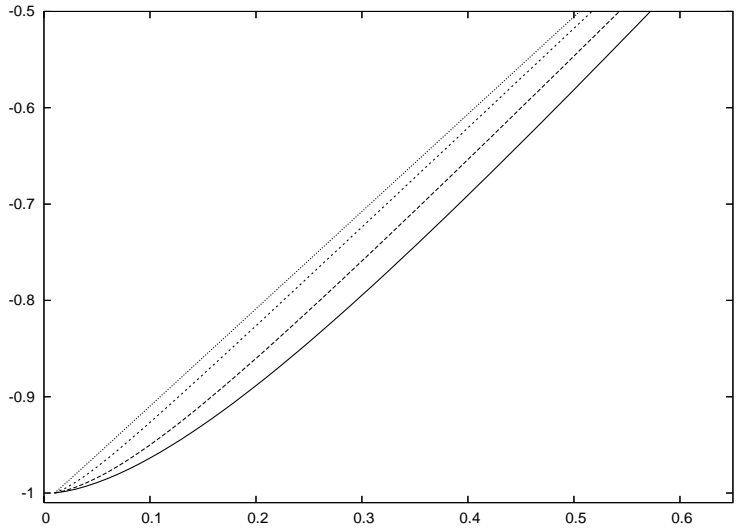}
\caption {\label{gamma1_red}
$\gamma^{(1)}(\sigma/L_x)/\abs{\gamma^{(1)}(0)}$ in terms of $\sigma/L_x$
for \mbox{$lq_M/(2\pi)$ = 1}; 2; 5 and 50 from bottom to top.
}
\end{center}
\end {figure}
\end{document}